\documentclass{iaus}
\usepackage{graphicx}
\usepackage{natbib}

\title[SFH of early-type galaxies at $z<1$] 
{Recent star formation in high-redshift early-type galaxies:
insights from the rest-frame UV}

\author[S. Kaviraj et al.]   
{S. Kaviraj$^1$\thanks{skaviraj@astro.ox.ac.uk}, S. K. Yi$^2$, E.
Gawiser$^3$, P. G. van Dokkum$^3$, S. Khochfar$^1$, K.
Schawinski$^1$ \and J. Silk$^1$}

\affiliation{$^1$Department of Physics, University of Oxford,
Keble Road, Oxford OX1 3RH, UK \\
$^2$Yonsei University, Centre for Space Astrophysics, Seoul
120749, Korea\\
$^{3}$Yale Center for Astronomy and Astrophysics, Yale University,
New Haven, CT 06520-8121}

\pubyear{2007}
\volume{245}  
\pagerange{}
\date{}
 \jname{IAU Symposium} \editors{M. Bureau,
E. Athanassoula, and B. Barbuy}
\begin{document}

\maketitle


\begin{abstract}
We combine deep $UBVRIzJK$ photometry from the MUSYC survey with
redshifts from the COMBO-17 survey to study the \emph{rest-frame}
ultraviolet ($UV$) properties of 674 high-redshift ($0.5<z<1$)
early-type galaxies, drawn from the Extended Chandra Deep Field
South (E-CDFS). Galaxy morphologies are determined through visual
inspection of Hubble Space Telescope (HST) images taken from the
GEMS survey. We harness the sensitivity of the $UV$ to young ($<1$
Gyrs old) stars to quantify the \emph{recent} star formation
history of the early-type population. We find compelling evidence
that early-types of all luminosities form stars over the lifetime
of the Universe, although the bulk of their star formation is
already complete at high redshift. Luminous ($-23<M(V)<-20.5$)
early-types form 10-15 percent of their mass after $z=1$, while
their less luminous ($M(V)>-20.5$) counterparts form 30-60 percent
of their mass in the same redshift range.

\end{abstract}


\firstsection 

\section{Introduction}
The vast majority of work on early-type galaxies in the past has
focussed on optical spectro-photometric data. A significant
drawback of optical photometry is its lack of sensitivity to
moderate amounts of recent star formation (RSF). While red optical
colours imply a high-redshift formation epoch for the \emph{bulk}
of the stellar population in early-type galaxies, the optical
spectrum remains largely unaffected by the minority of stellar
mass that forms in these systems at low and intermediate redshift.
As a result it is virtually impossible to quantify early-type star
formation histories (SFHs) over the last 8 billion years
($0<z<1$).

RSF can be efficiently traced using the rest-frame ultraviolet
($UV$) spectrum, which is sensitive to young, massive main
sequence stars with ages less than $\sim$1 Gyr. Using a large
sample of early-type galaxies detected by the GALEX $UV$ space
telescope, Kaviraj et al. (2006) have recently shown that,
contrary to the expectations of traditional `monolithic collapse'
models \citep[e.g.][]{BLE92}, local early-types show widespread
recent star formation (RSF) - at least 30 percent show blue
$UV$-optical colours indicative of unambiguous RSF in these
systems. The work presented here extends these results to high
redshift by exploiting deep optical photometry to trace the
rest-frame $UV$ properties of early-type galaxies in the redshift
range $0.5<z<1$.


\section{Estimation of recent star formation}
The RSF in individual galaxies is calculated by comparing their
multi-wavelength photometry with synthetic galaxy populations,
generated in the framework of the standard model. Synthetic
populations are generated using the semi-analytical model of
\citet{Khochfar2003}. A library of synthetic photometry is
constructed by combining each of $\sim$15,000 model galaxies with
a single metallicity in the range $0.1Z_{\odot}$ to
$2.5Z_{\odot}$, dust extinction parametrised by a value of
$E(B-V)$ in the range 0 to 0.5 and convolving them with the
stellar models of \citet{Yi2003}. We construct an `average SFH'
for each galaxy, by combining the SFHs of all models in the
library weighted by their individual likelihoods. The RSF, defined
as the \emph{mass fraction of stars formed within the last Gyr},
is calculated from this average SFH. Figure 1 shows the derived
rest-frame $(NUV-r)$ colour-magnitude relation and the values of
RSF in the E-CDFS galaxy population.


\section{Summary and discussion}

\begin{figure}
$\begin{array}{cc}
\includegraphics[width=3.3in]{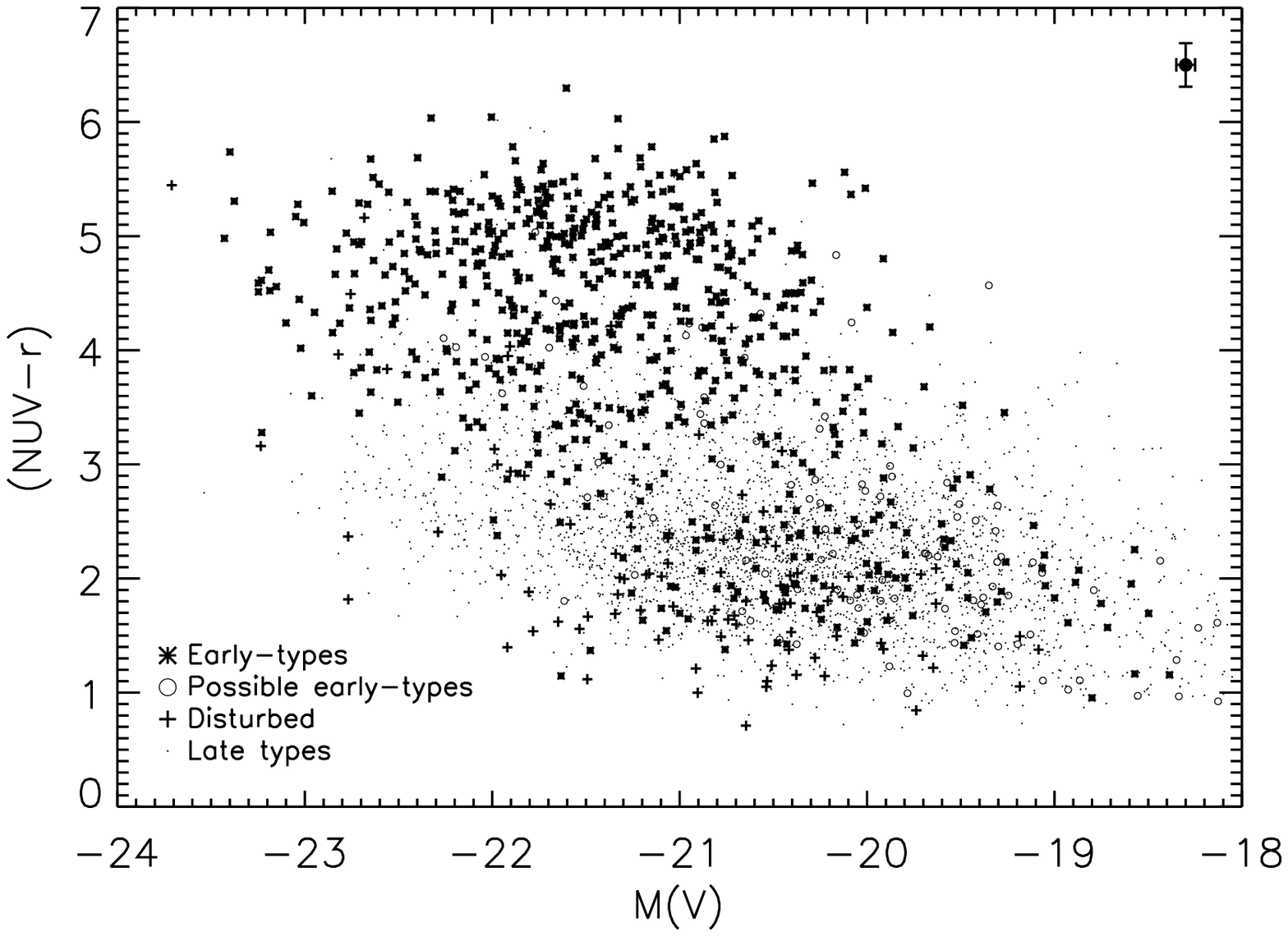} &
\includegraphics[width=3.3in]{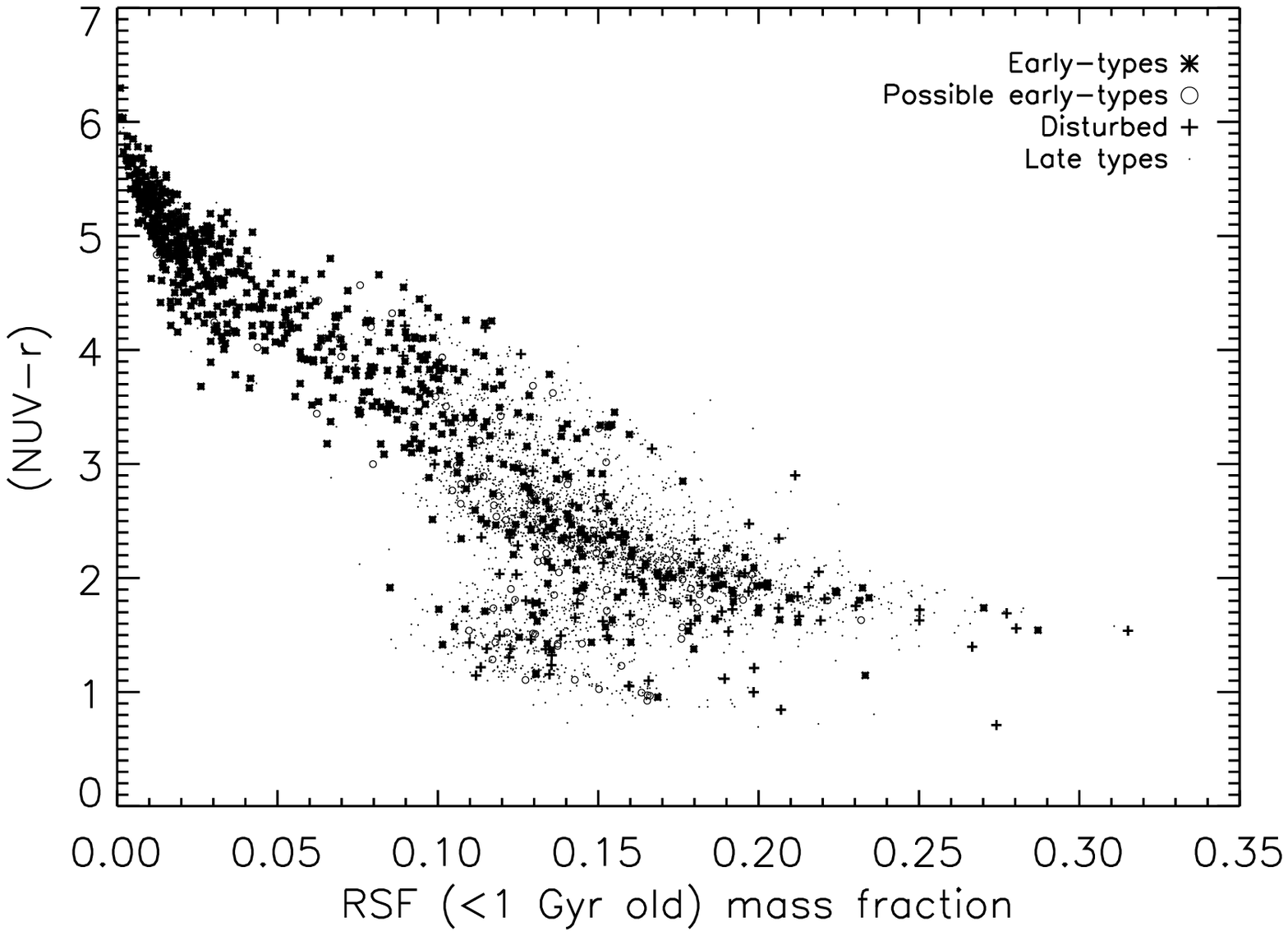}
\end{array}$
\caption{LEFT-HAND PANEL: Rest-frame $(NUV-r)$ colour-magniutude
relation of the E-CDFS galaxy population. RIGHT-HAND PANEL:
Rest-frame $(NUV-r)$ colour plotted against the RSF in the E-CDFS
galaxy population. The $NUV$ ($\sim2300\AA$) and $r$ band filters
are taken from the GALEX and SDSS filtersets respectively.}
\label{fig:discussion}
\end{figure}

The early-type population as a whole exhibits a typical RSF
between 5 and 13 percent in the redshift range $0.5<z<1$, while
the early-types on the broad `red sequence', ($NUV-r>4$),
typically show RSF values less than 5 percent. The reddest
early-types (which are also the most luminous) are virtually
quiescent with RSF values of $\sim1$ percent. Since the timescale
of this study is $\sim2.5$ Gyrs, a simple extrapolation (from RSF
values in Figure 1) indicates that luminous ($-23<M(V)<-20.5$)
early-type galaxies typically form up to 10-15 percent of
their mass after $z=1$ (with a tail to higher values), while less
luminous early-types ($M(V)>-20.5$) form 30-60 percent of their
mass after $z=1$.


\nocite{Kaviraj2006b}

\bibliographystyle{chicago}
\bibliography{references}


\end{document}